\documentclass[prl,twocolumn,showpacs]{revtex4}%
\usepackage{graphicx}

\begin{document}

\title{Nonequilibrium pattern formation in chiral Langmuir monolayers with transmembrane flows}
\author{Tatsuo Shibata$^{1}$ and Alexander S. Mikhailov$^{2}$}

\begin{abstract}
Nonequilibrium Langmuir monolayers including a fraction of chiral molecules
and subject to transmembrane flow are considered. The flow induces coherent
collective precession of chiral molecules. Our theoretical study shows that
splay interactions in this system lead to spatial redistribution of chiral
molecules and formation of spiral waves and target patterns observed in experiments.
\end{abstract}

\address{$^{1}$Department of  Mathematics and Life Sciences, Hiroshima
University,  1-3-1,  Kagamiyama, Higashi-Hiroshima, 739-8526, Japan,  and
$^{2}$Abteilung  Physikalische Chemie, Fritz-Haber-Institut der\\
Max-Planck-Gesellschaft, Faradayweg 4-6, 14195 Berlin, Germany}


\maketitle

Studies of pattern formation in nonequilibrium soft matter are essential for
understanding the operation of biological cells and for potential
applications~\cite{Mikhailov}. Biological membranes including active molecular
pumps or channels can develop shape oscillations and show persistent wave
propagation~\cite{Ramaswamy,Chen}. If a membrane contains rotating molecular
motors, interactions between them may lead to the development of regular
arrays with the crystalline order~\cite{Lenz}. Phase separation in
two-component lipid layers is responsible for budding and replication of
vesicles~\cite{Kumar,Takakura}. Closely related to biomembranes, Langmuir
monolayers are formed by organic lipid or amphiphilic molecules disposed on a
liquid-gas interface~\cite{Kaganer}. Nonequilibrium patterns of traveling
orientation waves in illuminated two-component Langmuir monolayers, where
illumination leads to transitions between different conformational states of
molecules, have been experimentally and theoretically
investigated~\cite{Tabe,Reigada,Reigada2,Okuzaono}. Recently, Tabe and
Yokoyama have demonstrated that Langmuir liquid-crystal monolayers including
chiral molecules (\textquotedblright molecular rotors\textquotedblright) are
easily brought to and maintained at nonequilibrium conditions by transmembrane
flows \cite{Tabe2}. If there is a gradient of small molecules across a
Langmuir monolayer, i.e. their concentrations in the liquid and the gas are
different, this produces a flow of such molecules through the monomolecular
layer. Such transmembrane flow gives rise to coherent collective precession of
molecular rotors. Experiments using reflected-light polarizing microscopy have
revealed that the precession is not uniform and complex orientational wave
patterns are observed. This behavior is apparently universal; it has been
verified for a number of chiral chemicals and different experimental
conditions \cite{Tabe2}. 

In this Letter, we construct a phenomenological theory of spatiotemporal
pattern formation in chiral Langmuir monolayers with transmembrane flows. An
important role in such systems is played by splay coupling between local
concentration of chiral molecules and the orientational field
~\cite{Selinger,Ohyama,Kamien}. We show that, in the presence of transmembrane
flow, this coupling gives rise to nonequilibrium wave patterns in the
orientational field and spatial redistribution of chiral molecules inside the
monolayer. The target patterns, seen in the experiments \cite{Tabe2}, should
thus be accompanied by aggregation of chiral molecules in the periphery of the
patterns. Other wave patterns, such as traveling stripes and rotating spiral
waves, are also possible. 

We study a model of an orientationally ordered two-component Langmuir
monolayer representing a mixture of chiral and achiral molecules (in the
experiments~\cite{Tabe2} the chiral molecules were making up only 10\% of the
monolayer). The local state of the monolayer is described by the variable
$c,$ giving the local fraction of chiral molecules, and by the director vector
$\mathbf{n}$ that represents the projection of the molecular tilt onto the
monolayer plane. The Landau free energy of the system is
\begin{eqnarray}
F &  =\int\bigl[\frac{1}{2}K(\nabla\mathbf{n})^{2}+k_{B}Tc\ln c+k_{B}%
T(1-c)\ln(1-c)\nonumber\\
&  +\frac{1}{2}G(\nabla c)^{2}+\Lambda c\nabla\cdot\mathbf{n}%
\bigr]dxdy.\label{eq.1}%
\end{eqnarray}
The first term corresponds to the elastic energy of orientational
ordering~($K$ is the Frank elastic constant). The next two terms determine the
lattice-gas entropy contribution to the free energy ($T$ is the temperature
and $k_{B}$ is the Boltzmann constant), and the following term (with the
coefficient $G$) takes into account weak energetic interactions between chiral
molecules which favor their uniform spatial distribution. The last term in the
expression for free energy describes splay interactions in the system. It
provides coupling between the scalar concentration field $c$ and the vector
orientational field $\mathbf{n}$~\cite{Selinger}; the parameter $\Lambda$
specifies the strength of splay interactions~\cite{Chiral}.

The kinetic equation for the local concentration $c$ of chiral molecules is
\begin{equation}
\dot{c}=\frac{D}{k_{B}T}\nabla\left[  c\left(  1-c\right)  \nabla
\frac{\delta F}{\delta c(\mathbf{r},t)}\right]  \label{eq.2}%
\end{equation}
where $D$ is their diffusion constant. The kinetic equations for the director
field $\mathbf{n}$ are
\begin{equation}
\dot{n}_{x}=-\Gamma\frac{\delta F}{\delta n_{x}(\mathbf{r},t)}+\Omega
n_{y},\quad\dot{n}_{y}=-\Gamma\frac{\delta F}{\delta n_{y}%
(\mathbf{r},t)}-\Omega n_{x}.\label{eq.3}%
\end{equation}
In addition to the relaxation terms ($\Gamma$ is the relaxation rate constant
for orientational ordering), we have phenomenologically included into these
equations, following Ref.\cite{Tabe2}, a term that describes planar precession
of the director vector. This precession is caused by the transmembrane flow
and its frequency $\Omega$ is linearly proportional to the flow intensity~(as
seen in the experiments \cite{Tabe2}). Because of the flow terms, the system
cannot relax to the state of thermal equilibrium and oscillations and active
wave propagation become possible.

Rescaling time and spatial coordinates as $t\rightarrow t(k_{B}T\Gamma)^{-1}$
and $\mathbf{r}\rightarrow\mathbf{r}\left(  K/k_{B}T\right)  ^{1/2}$ and using
the angle variable $\phi$ defined by $\mathbf{n}=(\cos\phi,\sin\phi),$ kinetic
equations (\ref{eq.2}) and (\ref{eq.3}) can be written in the form
\begin{equation}
\dot{c}=\nu\left[  \nabla^{2}c-g\nabla\left(  c\left(  1-c\right)
\nabla^{3}c\right)  +\lambda\nabla\left(  c(1-c)\nabla(\nabla\cdot
\mathbf{n}\right)  \right]  ,\label{eq.4}%
\end{equation}%
\begin{equation}
\dot{\phi}=\nabla^{2}\phi-\omega+\lambda\left(  \cos\phi\frac{\partial
c}{\partial y}-\sin\phi\frac{\partial c}{\partial x}\right)  .\label{eq.5}%
\end{equation}
The coefficients in these equations are $\nu=D(K\Gamma)^{-1}$, $g=GK^{-1}$,
$\lambda=\Lambda/(k_{B}TK)^{1/2}$ and $\omega=\Omega(k_{B}T\Gamma)^{-1}$. Note
that the total amount of chiral molecules is conserved and average spatial
concentration $c_{0}$ of these molecules is a parameter of the system.
According to equation~(\ref{eq.4}), splay coupling to the director field leads
to physical forces acting on chiral molecules and to the viscous flow of these
molecules in the monolayer plane. On the other hand, spatial gradients of
concentration $c$ lead, according to equation~(\ref{eq.5}), to local rotation
of the director vector $\mathbf{n}$.

When transmembrane flow is absent ($\omega=0$), these kinetic equations
describe relaxation to thermal equilibrium. The stationary equilibrium state
is uniform if splay interactions are sufficiently weak. If the splay
interaction strength $\lambda$ exceeds the critical value $\lambda
_{cr}=\left[  c_{0}(1-c_{0})\right]  ^{-1/2}$, the uniform state becomes
however unstable with respect to growth of spatial modes with the wavenumbers
$0<k<k_{\max}$, where $k_{\max}^{2}\propto(\lambda-\lambda_{cr})/g.$ This
instability has previously been investigated and is known to lead to the
formation of an equilibrium periodic stripe pattern~\cite{Selinger}. In this
equilibrium pattern, both the local concentration and the director orientation
are periodically varying along a certain direction.

To investigate nonequilibrium pattern formation induced by transmembrane flow,
numerical simulations of the model (\ref{eq.4}) and (\ref{eq.5}) using the
explicit Euler scheme with constant coordinate and time steps were performed.
In all simulations, periodic boundary conditions were applied.

\begin{figure}
\centering 
%\onefigure[width=.45\textwidth]{Fig1.eps}
\includegraphics[width=.45\textwidth]{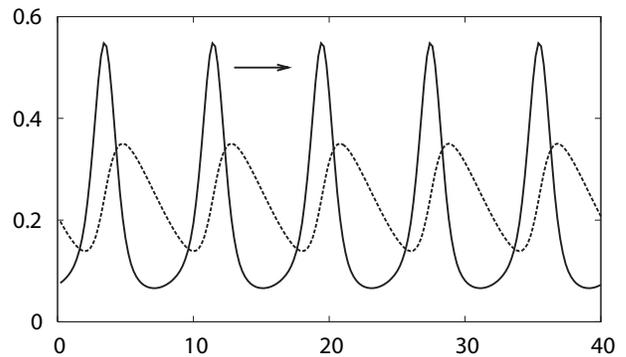}
\caption{Profiles of
concentration~$c$ (solid) and azimuthal angle~$\phi/2\pi$ (dashed) in a
traveling stripe pattern for $c_{0}=0.2,\nu=0.1,$ $g=1$, $\lambda=3,$ and
$\omega=0.005$. The arrow shows the direction of motion. }%
\label{fig:1}%
\end{figure}

When the transmembrane flow is introduced ($\omega\neq0$), we see that the
equilibrium stationary stripe pattern begins to move at a velocity that
increases with the flux intensity $\omega$. Figure 1 shows profiles of
concentration and azimuthal angle across a traveling stripe pattern. Note that
the shape of the stripes and their spatial period are not significantly
different from the respective equilibrium pattern at $\omega=0$..

\begin{figure}
\centering 
\includegraphics[width=.45\textwidth]{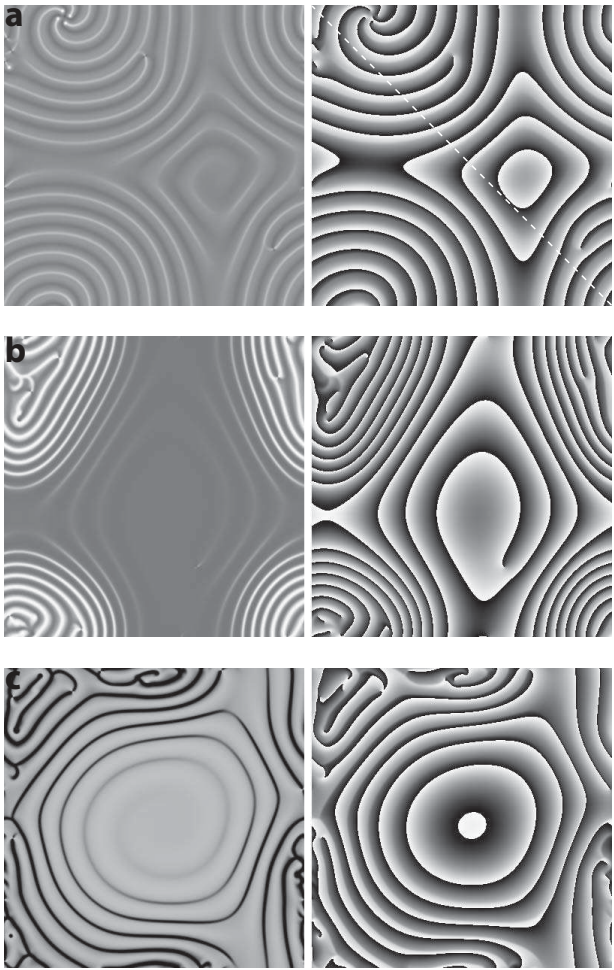}
\caption{Distributions
of concentration (left panels) and azimuthal angle (right panels) in
self-organized wave patterns obtained starting from random initial conditions
for systems with the parameters: (a) $c_{0}=0.1,\lambda=3.4,\omega
=0.01,\nu=10,g=10$ and (b) $c_{0}=0.1,$ $\lambda=3,$ $\omega=0.015$
$\nu=0.01,g=1$ and (c) $c_{0}=0.9,\lambda=3.4,\omega=0.015.,\nu=0.01,g=10,$
The linear size of the medium is $L=800$ (a) and $L=200$(b and c). The
concentration is displayed in gray scale with the darker color corresponding
to lower concentration values.}%
\label{fig:2}%
\end{figure}

If a simulation is started with random initial conditions for the azimuthal
angle field $\phi$, regular stripes are not formed. Instead, the system
undergoes relaxation to an equilibrium state with many spiral-shaped
orientational defects. Application of the transmembrane flow to a system in
this state leads, after a transient, to complex self-organized wave patterns.
Several examples of such patterns for different parameter values are shown in
Fig.\ref{fig:2} and movies~\cite{movie}.

\begin{figure}
\centering 
\includegraphics[width=.35\textwidth]{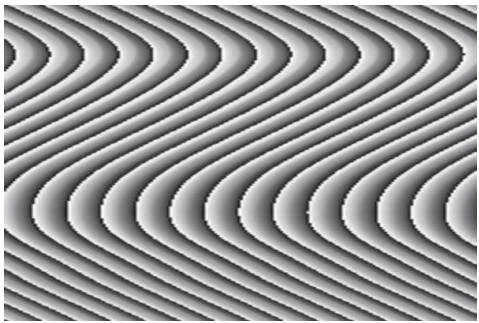}
%50*200=10000
\caption{Space-time diagrams displaying evolution of the azimuthal angle along
the diagonal cross section shown by dashed line in Fig.2a. Time runs from left
to right in the horizontal direction, the total shown time interval is
$T=10000$. The target pattern and the wave sink (above) are seen. }%
\label{fig:3}%
\end{figure}

The central region in the pattern shown in Fig.2a is periodically emitting
orientation waves. Repeated generation and outward propagation of these waves
is seen in the space-time diagram Fig.\ref{fig:3} which displays evolution of
the azimuthal angle distribution along the diagonal cross section indicated by
the dashed line in Fig.2a. The waves spread out in the large central region
and run into the periphery part of the pattern, occupied by stripes with a
shorter wavelength. The corners of the medium in Fig.2a are occupied by
spiral-shaped stripe structures (because periodic boundary conditions are
used, they represent four parts of the same compact pattern). This stripe
structure represents a wave sink, as can be seen from the space-time diagram
in Fig.\ref{fig:3} (the sink occupies the upper part of this diagram) and by
examining the respective movie \cite{movie}.
While a target pattern is seen in
the center for the azimuthal angle distributions (right panel), a spiral wave
occupying the central region is seen in the concentration distribution (left
panel in Fig. 2a).

\begin{figure}
\centering 
\includegraphics[width=.45\textwidth]{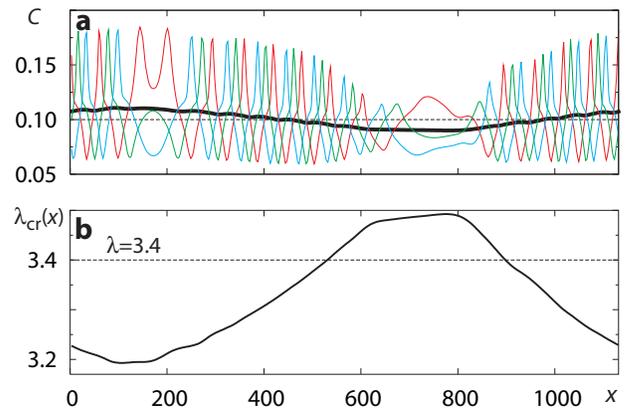}
\caption{(a)
Snapshots of concentration profiles $c(x,t)$ along the diagonal dashed line in
Fig.2a for three subsequent time moments (color online: red, green, blue); the
black solid curve shows the average smooth profile $\overline{c}(x).$ (b)
Effective local critical splay interaction $\lambda_{cr}(x)$ along the same
line.}%
\label{fig:4}%
\end{figure}

Figure 4a displays a superposition of three subsequent snapshots of
concentration profiles of chiral molecules along the diagonal line in Fig. 2a.
Additionally, we show here the concentration profile $\overline{c}(x)$
smoothed over the spatial scale $\Delta x=56$ (roughly the stripe period) and
averaged over the time interval of 10000. Although spatial variations are substantial, we
find that, on the average, the central region of the target pattern is
depleted of chiral molecules which become concentrated in the curled stripes
in its periphery.

The spatial redistribution of chiral molecules allows to qualitatively explain
the emergence of target-shaped wave patterns. When a stripe pattern, caused by
splay interactions, is formed, it tends to hinder the angular rotation forced
by the transmembrane flow. Therefore, oscillations develop only in the areas
free from the stripes. As we have seen above, the critical splay interaction
strength, needed for the formation of stripes, depends on the concentration as
$\lambda_{cr}(c)=\left[  c(1-c)\right]  ^{-1/2}$. If the concentration is not
constant and smoothly varies in space, the critical conditions will be
determined by the \textit{local} concentration. As seen from Fig.4a, the
concentration varies rather rapidly inside the stripe regions. If we smooth
the concentration profile and substitute $\overline{c}(x)$ instead of $c$ in
the expression for the critical splay interaction strength, we obtain the
effective dependence $\lambda_{cr}(x)$ displayed in Fig. 4b. For comparison,
the horizontal dash line shows the value of $\lambda$ in the respective
simulation. Thus, inside the central area we have $\lambda<\lambda_{cr}(x),$
which explains why stripes are absent here. The boundary where the stripes
first develop is roughly determined by the condition $\lambda<\lambda
_{cr}(x).$ In the region filled with stripes, $\lambda_{cr}(x)$ lies below
$\lambda,$ so that the uniform state is unstable. Of course, this argument is
only approximate because the stripes are not at equilibrium. However, the
effect of transmembrane flow on the stripe pattern is relatively weak, as we
have already noticed. 

The simulation shown in Fig. 2a has been performed assuming strong diffusion
of chiral molecules ($\nu=10$). When diffusion is weaker ($\nu=0.01$ in Fig.
2b), depletion of chiral molecules in the central region and their
accumulation inside the stripe structure in the periphery become strongly
pronounced. Now, the characteristic wavelength of the stripes in the periphery
of the target pattern is much shorter than the wavelength of the equilibrium
stripe pattern for~$c=c_{0}$. This is because the wavelength of the stripe
pattern decreases when the difference~$|\lambda-\lambda_{cr}(x)|$ becomes
larger. The stripes move only very slowly in the periphery under these
conditions. The large central region can contain rotating spiral waves, as
seen in Fig. 2b.The splay intensity strength $\lambda=3$ in this simulation
was below the critical strength $\lambda_{cr}=3.33\cdots$ corresponding to
$c_{0}=0.1.$ This means that the uniform state remains stable with respect to
small perturbations. However, strong initial perturbations can still lead in
this case to the formation of steady wave patterns. The spiral-shaped stripe
pattern in Fig. 2b is similar to the equilibrium spiral
patterns~\cite{Selinger2}. Note that a number of topological orientational
defects are present in the pattern shown in Fig. 2a, but all of them, except
one, belong to the dense stripe region in the periphery.

So far, only patterns for low average concentrations of chiral molecules have
been discussed. In contrast to this, the wave pattern shown in Fig. 2c
corresponds to a high concentration of chiral molecules ($c_{0}=0.9$).
Analyzing this pattern, several significant differences are seen. The
concentration of chiral molecules is now \textit{increased} inside the uniform
central region and decreased in the region occupied by the stripes. The
difference in the spatial distribution of chiral molecules for $c_{0}=0.9$ can
be explained if we notice that $\lambda_{cr}(c)=\left[  c(1-c)\right]
^{-1/2}$ depends non-monotonously on concentration $c$ and increases with
concentration when $c>0.5$. Therefore, inside the central region the formation
of stripes is prevented because the condition $\lambda<\lambda_{cr}(x)$ again
holds, now because the chiral molecules have aggregated in this region,
pushing the achiral component into the stripe-filled periphery region.

Thus, we have shown that splay interactions, based on coupling between the
orientational and concentration fields, determine principal properties of
nonequilibrium wave patterns in chiral Langmuir monolayers subject to the
transmembrane flow. The theory explains target-shaped and spiral wave patterns
observed in the experiments \cite{Tabe2,Tabe3}.  It relates the
formation of such patterns to nonequilibrium spatial redistribution of
molecular rotors, tending to aggregate in the areas occupied by slowly
traveling, densely packed stripes. 

Biomembranes are closely related to Langmuir monolayers and we expect that
similar results should hold, under appropriate conditions, also for the
membranes including a fraction of chiral molecules. The transmembrane flow in
such systems is created by a gradient of concentration of small molecules or
ions that leak through the membrane. The leakage may bring the membrane to
nonequilibrium conditions, giving rise to traveling waves and complex
self-organized wave patterns. Importantly, chiral molecules (and, possibly,
some passive inclusions) can then be transported and spatially redistributed
in a membrane as a result of wave propagation.

\acknowledgements

The authors acknowledge stimulating discussions with Y. Tabe and H. Yokoyama. Computer simulations were performed using a parallel supercomputer in the Yukawa Institute for Theoretical Physics. One of us (T.S.) acknowledges financial support through a grant for young scientists from the Ministry of Education, Culture, Sports, Science and Technology in Japan.


\begin{thebibliography}{99}
\bibitem {Chiral}If a monolayer consists of chiral molecules, the splay term
generally is $\Lambda c\nabla\cdot\mathbf{n}^{\prime}$ where $\mathbf{n}
^{\prime}$ is obtained from $\mathbf{n}$ by rotation by some fixed angle
$\alpha$. Since all other terms in the free energy are, however, invariant
with respect to such rotation, it is not important for pattern formation and
we put here $\alpha=0$.
\end{thebibliography}
\end{document}